\documentstyle[12pt]{article}
\def\hybrid{\topmargin -20pt	\oddsidemargin 0pt
	\headheight 0pt	\headsep 0pt
	\textwidth 6.25in	% A4 paper
	\textheight 9.5in	% A4 paper
	\marginparwidth .875in
	\parskip 5pt plus 1pt	\jot = 1.5ex}

\hybrid

\def\baselinestretch{1.2}

\catcode`\@=11

\def\marginnote#1{}
\newcount\hour
\newcount\minute
\newtoks\amorpm
\hour=\time\divide\hour by60
\minute=\time{\multiply\hour by60 \global\advance\minute by-\hour}
\edef\standardtime{{\ifnum\hour<12 \global\amorpm={am}%
	\else\global\amorpm={pm}\advance\hour by-12 \fi
	\ifnum\hour=0 \hour=12 \fi
	\number\hour:\ifnum\minute<10 0\fi\number\minute\the\amorpm}}
\edef\militarytime{\number\hour:\ifnum\minute<10 0\fi\number\minute}
\def\draftlabel#1{{\@bsphack\if@filesw {\let\thepage\relax
   \xdef\@gtempa{\write\@auxout{\string
      \newlabel{#1}{{\@currentlabel}{\thepage}}}}}\@gtempa
   \if@nobreak \ifvmode\nobreak\fi\fi\fi\@esphack}
	\gdef\@eqnlabel{#1}}
\def\@eqnlabel{}
\def\@vacuum{}
\def\draftmarginnote#1{\marginpar{\raggedright\scriptsize\tt#1}}

\def\draft{\oddsidemargin -.5truein
	\def\@oddfoot{\sl preliminary draft \hfil
	\rm\thepage\hfil\sl\today\quad\militarytime}
	\let\@evenfoot\@oddfoot	\overfullrule 3pt
	\let\label=\draftlabel
	\let\marginnote=\draftmarginnote
   \def\@eqnnum{(\theequation)\rlap{\kern\marginparsep\tt\@eqnlabel}%
\global\let\@eqnlabel\@vacuum}  }

\def\preprint{\twocolumn\sloppy\flushbottom\parindent 2em
	\leftmargini 2em\leftmarginv .5em\leftmarginvi .5em
	\oddsidemargin -.5in	\evensidemargin -.5in
	\columnsep .4in	\footheight 0pt
	\textwidth 10.in	\topmargin  -.4in
	\headheight 12pt \topskip .4in
	\textheight 6.9in \footskip 0pt
	\def\@oddhead{\thepage\hfil\addtocounter{page}{1}\thepage}
	\let\@evenhead\@oddhead	\def\@oddfoot{}	\def\@evenfoot{} }

\def\numberbysection{\@addtoreset{equation}{section}
	\def\theequation{\thesection.\arabic{equation}}}

\def\underline#1{\relax\ifmmode\@@underline#1\else
	$\@@underline{\hbox{#1}}$\relax\fi}

\def\titlepage{\@restonecolfalse\if@twocolumn\@restonecoltrue\onecolumn
     \else \newpage \fi \thispagestyle{empty}\c@page\z@
	\def\thefootnote{\fnsymbol{footnote}} }

\def\endtitlepage{\if@restonecol\twocolumn \else \newpage \fi
	\def\thefootnote{\arabic{footnote}}
	\setcounter{footnote}{0}}  %\c@footnote\z@ }

\catcode`@=12
\relax

\def\figcap{\section*{Figure Captions\markboth
	{FIGURECAPTIONS}{FIGURECAPTIONS}}\list
	{Figure \arabic{enumi}:\hfill}{\settowidth\labelwidth{Figure
999:}
	\leftmargin\labelwidth
	\advance\leftmargin\labelsep\usecounter{enumi}}}
 \relax
\def\tablecap{\section*{Table Captions\markboth
	{TABLECAPTIONS}{TABLECAPTIONS}}\list
	{Table \arabic{enumi}:\hfill}{\settowidth\labelwidth{Table
999:}
	\leftmargin\labelwidth
	\advance\leftmargin\labelsep\usecounter{enumi}}}
 \relax
\def\reflist{\section*{References\markboth
	{REFLIST}{REFLIST}}\list
	{[\arabic{enumi}]\hfill}{\settowidth\labelwidth{[999]}
	\leftmargin\labelwidth
	\advance\leftmargin\labelsep\usecounter{enumi}}}
 \relax
\makeatletter
\newcounter{pubctr}
\def\publist{\@ifnextchar[{\@publist}{\@@publist}}
\def\@publist[#1]{\list
	{[\arabic{pubctr}]\hfill}{\settowidth\labelwidth{[999]}
	\leftmargin\labelwidth
	\advance\leftmargin\labelsep
	\@nmbrlisttrue\def\@listctr{pubctr}
	\setcounter{pubctr}{#1}\addtocounter{pubctr}{-1}}}
\def\@@publist{\list
	{[\arabic{pubctr}]\hfill}{\settowidth\labelwidth{[999]}
	\leftmargin\labelwidth
	\advance\leftmargin\labelsep
	\@nmbrlisttrue\def\@listctr{pubctr}}}
 \relax
\makeatother
\newskip\humongous \humongous=0pt plus 1000pt minus 1000pt

\newif\ifdtup

\relax

\def\be{\begin{equation}}
\def\ee{\end{equation}}
\def\ba{\begin{eqnarray}}
\def\ea{\end{eqnarray}}

\def\del{\partial}

\def\r{\rho}
\def\a{\alpha}

\def\b{\beta}

\def\d{\delta}

\def\th{\theta}

\def\m{\mu}
\def\n{\nu}
\def\om{\omega}
\def\Om{\Omega}
\def\l{\lambda}
\def\L{\Lambda}
\def\s{\sigma}

\def\bs{\bigskip}
\def\no{\noindent}

\def\IR{\relax{\rm I\kern-.18em R}}

\def \ha {{1\over 2}}

\def \ov {\over}

\def\IR{\relax{\rm I\kern-.18em R}}

\def\IR{\relax{\rm I\kern-.18em R}}

\begin{document}
%\draft

%\renewcommand{\theequation}{\thesection.\arabic{equation}}
\renewcommand{\theequation}{\arabic{equation}}
\newcommand{\beq}{\begin{equation}}
\newcommand{\eeq}[1]{\label{#1}\end{equation}}
\newcommand{\ber}{\begin{eqnarray}}
\newcommand{\eer}[1]{\label{#1}\end{eqnarray}}
\newcommand{\eqn}[1]{(\ref{#1})}
\begin{titlepage}
\begin{center}

\hfill THU--96/33\\
\hfill September 1996\\
\hfill hep--th/9609165\\

\vskip .8in

{\large \bf COSET MODELS AND DIFFERENTIAL GEOMETRY
\footnote{Contribution to the proceedings of the 
{\em Conference on Gauge Theories, 
Applied Supersymmetry and Quantum Gravity}, Imperial College, London, 
5-10 July 1996 and the e--proceedings 
of Summer 96 Theory Institute, {\em Topics in Non-Abelian Duality},
Argonne, IL, 27 June - 12 July 1996. }}

\vskip 0.6in

{\bf Konstadinos Sfetsos
\footnote{e--mail address: sfetsos@fys.ruu.nl}}\\
\vskip .1in

{\em Institute for Theoretical Physics, Utrecht University\\
     Princetonplein 5, TA 3508, The Netherlands}\\

\vskip .2in

\end{center}

\vskip .6in

\begin{center} {\bf ABSTRACT } \end{center}
\begin{quotation}\noindent

\no
String propagation on a
curved background defines an embedding problem of surfaces 
in differential geometry.
Using this, we show that in a wide class of backgrounds the 
classical dynamics of the physical degrees of freedom of the string 
involves 2--dim
$\s$--models corresponding to coset conformal field theories.

\vskip .2in

\noindent

\end{quotation}

\end{titlepage}
%\vfill
%\eject

\def\baselinestretch{1.2}
\baselineskip 16 pt
\noindent

Coset models have been used in string theory 
for the construction of classical vacua,
either as internal theories in string compactification or as
exact conformal field theories representing curved spacetimes.
Our primary aim in this note, based on \cite{basfe}, is to reveal their 
usefulness in a different context by
demonstrating that certain 
classical aspects of constraint systems are governed by 2--dim 
$\s$--models corresponding to some specific
coset conformal field theories.
In particular, we will examine string propagation
on arbitrary curved backgrounds with Lorentzian signature which 
defines an embedding problem in differential geometry,
as it was first shown for 4--dim Minkowski space by Lund and 
Regge \cite{LuRe}.
Choosing, whenever possible, the temporal gauge one may solve the Virasoro
constraints and hence be left with $D-2$ 
coupled non--linear differential equations governing the dynamics of
the physical degrees of freedom of the string.
By exploring their integrability properties, and considering 
as our Lorentzian
background $D$--dim Minkowski space or the product form
$R\otimes K_{D-1}$, where $K_{D-1}$ is any WZW model
for a semi--simple compact group,
we will establish 
connection with the coset model conformal field theories
$SO(D-1)/SO(D-2)$.
This universal behavior irrespectively of the particular WZW model
$K_{D-1}$ is rather remarkable, 
and sheds
new light into the differential geometry of embedding surfaces using
concepts and field variables, which so far have been natural 
only in conformal field theory.

Let us consider classical propagation of closed strings on a 
$D$--dim background that is 
the direct product of the real line $R$ (contributing a minus 
in the signature matrix)
and a general manifold (with Euclidean signature) $K_{D-1}$.
We will denote $\s^\pm= \ha(\tau\pm \s)$, where
$\tau$ and $\s$ are the natural time and spatial variables 
on the world--sheet $\Sigma$.
Then,
the 2--dim $\s$--model action is given by
\be
S= \ha \int_\Sigma (G_{\m\n} + B_{\m\n}) \del_+ y^\m \del_- y^\n 
- \del_+ y^0 \del_- y^0 ~ , ~~~~~~~~ \m,\n =1,\dots , D-1~ ,
\label{smoac}
\ee
where $G$, $B$ are the non--trivial metric 
and antisymmetric tensor fields and
are independent of $y^0$.
The conformal gauge, we have implicitly chosen in writing 
down (\ref{smoac}),
allows us to further set $y^0=\tau$ (temporal gauge). 
Then we are left with the $D-1$ equations
of motion corresponding to the $y^\m$'s,
as well as with the Virasoro constraints 
\be
G_{\m\n} \del_\pm y^\m \del_\pm y^\n = 1 ~ ,
\label{cooss}
\ee
which can be used to further reduce 
the degrees of freedom by one, thus leaving only the 
$D-2$ physical ones.
We also define an angular variable $\th$ via the relation
\be
G_{\m\n} \del_+ y^\m \del_- y^\n = \cos \th ~ .
\label{angu}
\ee

In the temporal gauge we may restrict our analysis
entirely on $K_{D-1}$ and on 
the projection of the string world--sheet $\Sigma$ on the
$y^0=\tau$ hyperplane. The resulting 2--dim surface
$S$ has Euclidean signature with metric given by 
the metric $G_{\m\n}$ on $K_{D-1}$ restricted on $S$.
Using (\ref{cooss}), (\ref{angu}) we find that the 
corresponding line element reads
\be
ds^2 = d{\s^+}^2 + d{\s^-}^2 + 2 \cos\th d\s^+ d\s^- ~ .
\label{dsS2}
\ee
In general, determining the classical
evolution of the string is equivalent to 
the problem of determining the 2--dim surface
that it forms as it moves.
Phrased in purely geometrical terms this is equivalent, 
in our case, to 
the embedding problem of the 2--dim surface $S$
with metric (\ref{dsS2}) into the $(D-1)$--dim space
$K_{D-1}$. The solution requires that a complete set 
of $D-1$ vectors tangent and normal to the surface $S$ as functions
of $\s_+$ and $\s_-$ is found. 
In our case the 2 natural tangent vectors are 
$\{\del_+ y^\m, \del_- y^\m\}$, 
whereas the remaining $D-3$ normal ones will be denoted by
$\{\xi^\m_\s, \s=3,4,\dots, D-1\}$. 
These vectors obey first order partial 
differential equations \cite{Eisenhart} that depend, as expected, 
on the detailed structure of 
$K_{D-1}$. Since we are only interested
in some universal aspects we will solely restrict to the
corresponding \underline{compatibility} equations. 
In general, these involve the 
Riemann curvatures for the metrics of the two spaces  
$S$ and $K_{D-1}$, as well as the second 
fundamental form with components $\Om^\s_{\pm\pm}$, 
$\Om^\s_{+-}=\Om^\s_{-+}$ and the third 
fundamental form ($\equiv$ torsion) with components 
$\m^{\s\tau}_\pm =-\m^{\tau\s}_\pm$ \cite{Eisenhart}. It turns out 
that the $D-1$ classical equations of motion for \eqn{smoac}
(in the gauge $y^0 = \tau$) and the two
constraints (\ref{cooss}) completely determine the components
of the second fundamental form $\Om^\s_{+-}$ \cite{basfe}.
In what follows we will also use instead of $\m_\pm^{\s\tau}$
a modified, by a term that
involves $H_{\m\n\r}=\del_{[\m}B_{\n\r]}$,
torsion $M_\pm^{\s\tau}$ \cite{basfe}.
Then the compatibility equations 
for the remaining components $\Om^\s_{\pm\pm}$ and
$M_\pm^{\s\tau}$ are \cite{basfe}:
\ba
&& \Om^\tau_{++} \Om^\tau_{--} + \sin\th \del_+ \del_- \th
 =  - R^+_{\m\n\a\b} 
\del_+ y^\m \del_+ y^\a \del_- y^\n \del_- y^\b ~ ,
\label{gc1} \\
&& \del_{\mp} \Om^\s_{\pm\pm} - M_\mp^{\tau\s} \Om^\tau_{\pm\pm} 
-{1\ov \sin\th} \del_\pm\th \Om^\s_{\mp\mp}
 =  R^\mp_{\m\n\a\b} 
\del_\pm y^\m \del_\pm y^\a \del_\mp y^\b \xi^\n_\s ~ ,
\label{gc2} \\
&& \del_+ M_-^{\s\tau} - \del_- M_+^{\s\tau} 
- M_-^{\r[\s} M_+^{\tau]\r} 
+ {\cos\th \ov \sin^2\th} \Om^{[\s}_{++} \Om^{\tau]}_{--} 
= R^-_{\m [\b \a]\n}
\del_+ y^\m \del_- y^\n \xi^\a_\s \xi^\b_\tau ~ ,
\label{gc3}
\ea
where the curvature tensors and the covariant derivatives $D^\pm_\m$ 
are defined using the generalized 
connections that include the string torsion
$H_{\m\n\r}$.\footnote{We have written \eqn{gc3}
in a slightly different form compared to the same equation in \cite{basfe}
using the identity $D^-_\m H_{\n\a\b} = R^-_{\m[\n\a\b]}$.}
Equations (\ref{gc1})--\eqn{gc3} 
are generalizations of the 
Gauss--Codazzi and Ricci equations for a surface
immersed in Euclidean space.
For $D\geq 5$ there are
$\ha (D-3)(D-4)$ more unknown functions ($\th$, $\Om^\s_{\pm\pm}$ 
and $M_\pm^{\s\tau}$) than equations in \eqn{gc1}--\eqn{gc3}.
However, there is an underlying gauge invariance \cite{basfe}
which accounts for the extra (gauge) degrees of freedom 
and can be used to eliminate them (gauge fix).

Making further progress with 
the embedding system of equations (\ref{gc1})--(\ref{gc3}) 
as it stands seems a difficult task. This is 
due to the presence of terms
depending explicitly on $\del_\pm y^\m$ and $\xi^\m_\s$,
which can only be determined by solving the
actual string evolution equations.
Moreover, a Lagrangian from which 
(\ref{gc1})--(\ref{gc3}) can be derived as equations of
motion is also lacking. Having such a description is advantageous in 
determining the operator content of the theory and for quantization.
Rather remarkably, all of these problems can be simultaneously
solved by considering for $K_{D-1}$ either flat space with zero torsion or
any WZW model based on a 
semi--simple compact group $G$, with $\dim(G)=D-1$. 
This is due to the identity 
\be
R^\pm_{\m\n\a\b} =  0 ~ ,
\label{rdho}
\ee
which is valid not only for flat space with zero torsion but also
for all WZW models \cite{zachos}.
Then we completely get rid of the bothersome terms on the right
hand side of (\ref{gc1})--(\ref{gc3}).\footnote{Actually, the same 
result is obtained by demanding the weaker condition 
$R^-_{\m\n\a\b}=R^-_{\m\a\n\b}$, but we are not aware of any examples
where these weaker conditions hold.}
In is convenient to
extend the range of definition of 
$\Om^\s_{++}$ and $M_\pm^{\s\tau}$ by appending new components
defined as: $\Om^2_{++}= \del_+ \th$, 
$M_+^{\s 2}= \cot \th \Om^\s_{++}$ and
$M_-^{\s2} = - \Om^\s_{--}/\sin\th$.
Then equations (\ref{gc1})--(\ref{gc3}) can be recast into the
suggestive form 
\ba
&& \del_- \Om^i_{++} + M_-^{ij} \Om^j_{++} = 0 ~ , 
\label{new1} \\
&& \del_+ M_-^{ij} - \del_- M_+^{ij} + [M_+,M_-]^{ij} = 0 ~ ,
\label{new2}
\ea
where the new index $i=(2,\s)$. 
Equation (\ref{new2}) is a 
zero curvature condition for the matrices $M_\pm$ and it is locally
solved by $M_\pm = \L^{-1} \del_\pm \L$, 
where $\L \in SO(D-2)$. Then (\ref{new1}) can be cast into
equations for $Y^i=\L^{i2} \sin \th$ \cite{basfe}
\be
\del_- \left( {\del_+ Y^i \ov \sqrt{1-\vec Y^2}} \right) = 0~ , 
~~~~~ i = 2,3,\dots ,D-1 ~ .
\label{fiin}
\ee
These equations were derived before in \cite{barba}, while 
describing
the dynamics of a free string propagating in $D$--dimensional
{\it flat} space--time. It is remarkable that they remain 
unchanged even if the flat $(D-1)$--dim space--like part is replaced
by a curved background corresponding to a general WZW model.
Nevertheless, it should be emphasized that 
the actual evolution equations of the normal and tangent
vectors to the surface are certainly different from those 
of the flat space free string and can be found in \cite{basfe}.

As we have already mentioned, it would be advantageous if 
(\ref{fiin}) (or an equivalent system) could be derived 
as classical equations of motion for a 2--dim action of the 
form 
\be
S = {1\ov 2\pi \a'} \int (g_{ij} + b_{ij}) 
\del_+ x^i \del_- x^j ~ , ~~~~~~~~ i,j = 1,2,\dots, D-2 ~ .
\label{dynsm}
\ee 
The above action has a $(D-2)$--dim target space and only 
models the non--trivial dynamics of the physical degrees 
of freedom of the
string which itself 
propagates on the background corresponding to \eqn{smoac} which
has a $D$--dim target space.
The construction of such an action involves
a non--local change 
of variables and is based on 
the observation \cite{basfe} that (\ref{fiin}) 
imply chiral conservation laws, which 
are the same as the 
equations obeyed by the classical
parafermions for the coset model $SO(D-1)/SO(D-2)$ \cite{BSthree}.

We recall that the classical $\s$--model action 
corresponding to a coset $G/H$ is derived from the associated
gauged WZW model and the result is given by
\be 
S= I_0(g) + {1\ov \pi \a'} \int
{\rm Tr}(t^a g^{-1} \del_+  g) M^{-1}_{ab} {\rm Tr}
(t^a  \del_- g g^{-1}) ~ , ~~~~
M^{ab} \equiv {\rm Tr}(t^a g t^b g^{-1}- t^a t^b) ~ ,
\label{dualsmo}
\ee
where $I_0(g)$ is the WZW action for a group element $g\in G$ and
$\{t^A\}$ are representation matrices of the Lie algebra for 
$G$ with indices split as $A=(a,\a)$, where $a\in H$ 
and $\a\in G/H$.
We have also assumed that a unitary gauge has been chosen 
by fixing $\dim(H)$ 
variables among the total number of $\dim(G)$ parameters
of the group element $g$. Hence, there are
$\dim(G/H)$ remaining variables, which will be denoted by $x^i$. 
The natural objects generating infinite dimensional symmetries
in the background \eqn{dualsmo} are the classical parafermions 
(we restrict to one chiral sector only) defined in general as \cite{BCR} 
\be
\Psi_+^\a = {i \ov \pi \a'} {\rm Tr} (t^\a f^{-1} \del_+ f ) ~ ,
~~~~~~~~~ f\equiv h_+^{-1} g h_+ \in G ~ ,
\label{paraf}
\ee
and obeying on shell $\del_- \Psi_+^\a = 0 $.
The group element $h_+\in H$ is given as a path order exponential 
using the on shell value of the gauge field $A_+$
\be
h_+^{-1} = {\rm P} e^{- \int^{\s^+} A_+}~ , ~~~~~~~~
A_+^a  =   M^{-1}_{ba}  {\rm Tr} (t^b g^{-1}\del_+ g) ~ .
\label{hphm}
\ee

Next we specialize to the $SO(D-1)/SO(D-2)$ gauged WZW models. 
In this case the index $a=(ij)$ and the index $\a=(0i)$ with
$i=1,2,\dots , D-2$. Then the 
parafermions \eqn{paraf} assume the 
form (we drop $+$ as a subscript) \cite{BSthree,basfe}
\ba
&& \Psi^i  = {i\ov \pi \a'}
{\del_+ Y^i\ov \sqrt{1-\vec Y^2}}  =  
  {i \ov \pi \a'} {1\ov \sqrt{1-\vec X^2}} (D_+X)^j  h_+^{ji} ~ ,
\nonumber \\
&& (D_+X)^j =  \del_+ X^j - A_+^{jk} X^k ~ , ~~~~~~ 
Y^i = X^j (h_+)^{ji}~ .
\label{equff}
\ea
Thus, equation $\del_- \Psi^i = 0$ is 
precisely (\ref{fiin}), whereas \eqn{dualsmo}
provides the action \eqn{dynsm} to our embedding problem.
The relation between the $X^i$'s and the $Y^i$'s in \eqn{equff}
provides 
the necessary non--local change of variables that transforms 
(\ref{fiin}) into a Lagrangian system of equations.
It is highly non--intuitive 
in differential geometry, and only 
the correspondence with parafermions makes it natural.

%\no
%The operator product expansion of two parafermions to next 
%to the leading order in the $\a'$--expansion is 
%equivalent to the corresponding Poisson bracket which were 
%computed for any $G/H$ coset model in \cite{BCR}. In our case 
%it reads
%\be
% \Psi^i(z) \Psi^j(w) =  {\d_{ij}/\a'\ov (z-w)^2} - {\a'\ov 2} \ln(z-w) \left(
%\d_{ij} \Psi(z) \cdot \Psi(w) - \Psi^j(z) \Psi^i(w) \right) + O(\a')^2 ~ .
%\label{poi}
%\ee
%Hence our parafermions would have been ordinary spin one bosonic currents
%if it wasn't for the last term in (\ref{poi}) which is responsible 
%for their non--trivial
%monodromy and braiding properties.

It remains to conveniently parametrize the group element 
$g\in SO(D-1)$. In the right coset decomposition with respect to 
the subgroup $SO(D-2)$ we may write \cite{BSthree}
\be
g = \left( \begin{array} {cc}
1 & 0 \\
  &  \\
0 & h  \\
\end{array}
\right) \cdot 
 \left( \begin{array} {cc}
b & X^j \\
  &  \\
- X^i  & \d_{ij} - {1\ov b+1} X^i X^j \\
\end{array}\right) ~ ,
\label{H} 
\ee
where $h\in SO(D-2)$ and $b \equiv \sqrt{1-\vec X^2}$.
The range of the parameters in the vector $\vec X$ is restricted 
by $\vec X^2\leq 1$.
A proper gauge fixing is to choose  
the group element $h$ in the Cartan torus of $SO(D-2)$ 
and then use the remaining gauge symmetry to gauge fix 
some of the components of the vector $\vec X$. 
If \underline{$D=2 N + 3= {\rm odd}$} then we may 
cast the orthogonal matrix $h\in SO(2N+1)$ and
the row vector $\vec X$ into the form \cite{basfe}
\ba
&& h={\rm diagonal}\left(h_1,h_2,\dots,h_N,1\right)~ ,~~~~~
h_i = \pmatrix{
\cos 2\phi_i & \sin 2\phi_i \cr
-\sin 2\phi_i & \cos 2\phi_i \cr} \nonumber \\ 
&& \vec X =\left(0,X_2,0,X_4,\dots,0,X_{2N},X_{2N+1}\right) ~ .
\label{hdixn}
\ea
On the other hand if \underline{$D=2 N + 2= {\rm even}$}
then $h\in SO(2N)$ can be gauge fixed in a  
form similar to the one in \eqn{hdixn} with the 1 removed. 
Similarly in the vector $\vec X$ there is no 
$X_{2N+1}$ component.
In both cases the total number of
independent variables is $D-2$, as it should be. 

\underline{\em Examples:}
As a first example we consider the Abelian coset $SO(3)/SO(2)$ \cite{BCR}. 
In terms of our original problem it arises after solving the 
Virasoro constraints for
strings propagating on 4--dim Minkowski space or on the 
direct product of the real line $R$ and the WZW model for $SU(2)$.
Using $X_2= \sin 2\th$ one finds that 
\be
A_+ = \pmatrix{ 0 & 1\cr -1 & 0 } 
(1- \cot^2\th) \del_+ \phi ~ ,
\label{gsu2}
\ee
and that the background corresponding to \eqn{dynsm}
has metric \cite{BCR} 
\be 
ds^2 = d\th^2  + \cot^2\th d\phi^2 ~ .
\label{S1}
\ee 
Using (\ref{equff}), the corresponding Abelian parafermions 
$\Psi_\pm = \Psi_2 \pm i\Psi_1$ assume the familiar form 
\be
\Psi_\pm = (\del_+ \th \pm i \cot\th \del_+ \phi) 
e^{\mp i \phi \pm i  \int \cot^2\th \del_+ \phi } ~ ,
\label{pasu2}
\ee
up to an overall normalization. An alternative way of seeing the 
emergence of the coset 
$SO(3)/SO(2)$ is from the original system of 
embedding equations \eqn{gc1}--\eqn{gc3} for $D=4$ and zero 
curvatures. They just reduce to the classical equations of 
motion for the 2--dim $\s$--model corresponding to the metric 
\eqn{S1} \cite{LuRe}, as it was observed in \cite{Baso3}.

Our second example is the simplest 
non--Abelian coset $SO(4)/SO(3)$ \cite{BSthree}. 
In our context it
arises in string
propagation on 5--dim Minkowski space or on the direct 
product of the real line $R$ and the WZW model based on 
$SU(2)\otimes U(1)$.
Parametrizing $X_2 = \sin 2\th \cos \om$
and $X_3 = \sin 2\th \sin \om$ one finds that 
the $3 \times 3$ antisymmetric matrix for the $SO(3)$ 
gauge field $A_+$ has independent components given by
\ba
A^{12}_+  & = & -\left( {\cos 2\th \ov \sin^2\th \cos^2\om }
+ \tan^2\om {\cos^2\th -\cos^2\phi \cos 2\th\ov 
\cos^2\th \sin^2 \phi} \right)
\del_+\phi - \cot\phi \tan\om \tan^2 \th 
\del_+\om ~ ,\nonumber \\ 
A^{13}_+ & = &  \tan\om
{\cos^2\th -\cos^2\phi \cos 2\th\ov 
\cos^2\th \sin^2 \phi} \del_+\phi
+ \cot\phi \tan^2 \th \del_+\om ~ ,
\label{expap} \\
A^{23}_+ & = & \cot\phi \tan \om {\cos 2\th\ov \cos^2\th} 
\del_+ \phi - \tan^2 \th \del_+\om ~ .
\nonumber 
\ea
Then, the 
background metric for the action \eqn{dynsm} governing the
dynamics of the 3 physical string degrees of freedom 
is \cite{BSthree}
\be
ds^2 = d\th^2 + \tan^2\th (d\om + \tan\om \cot \phi d\phi)^2 
+ {\cot^2\th \ov \cos^2\om} d\phi^2 ~ ,
\label{ds3}
\ee
and the antisymmetric tensor is zero.
The parafermions of the $SO(4)/SO(3)$ coset are non--Abelian and are
given by (\ref{equff}) with some explicit expressions 
for the covariant derivatives \cite{basfe}.
In addition to the two examples above, there also
exist explicit results for the coset $SO(5)/SO(4)$
\cite{BShet}.
This would correspond in our context to string
propagation on a 6--dim Minkowski space or 
on the background 
$R$ times the $SU(2)\otimes U(1)^2$
WZW model.

An obvious extension one could make is to 
consider the same embedding problem but with Lorenzian instead of
Euclidean backgrounds representing the ``spatial'' part $K_{D-1}$.
This would necessarily involve $\s$--models for cosets based on
non--compact groups. The case for $D=4$
has been considered in \cite{vega}.
It is interesting to consider supersymmetric
extensions of the present work in connection also with \cite{susyre}.
In addition, formulating
classical propagation of $p$--branes
on curved backgrounds as a geometrical problem of embedding surfaces 
(for work in this direction see \cite{kar}) and 
finding the $p+1$--dim $\s$--model action (analog of \eqn{dynsm} for 
strings ($p=1$)) that governs the 
dynamics of the physical degrees of freedom of the $p$--brane 
is an open interesting problem.

The techniques we have presented in this note can also be used to 
find the Lagrangian description of the symmetric space 
sine--Gordon models \cite{Pohlalloi} which 
have been described as perturbations of coset conformal field
theories \cite{bapa}.
Hence, the corresponding parafermion variables will play
the key role in such a construction.

Finally, an interesting issue is the quantization of constrained 
systems.
Quantization in string theory usually proceeds by quantizing 
the unconstrained degrees of freedom and then imposing the 
Virasoro constraints
as quantum conditions on the physical states. 
However, in the
present framework the physical degrees of freedom should be quantized
directly using the quantization of the associated parafermions.
Quantization of the $SO(3)/SO(2)$ parafermions has been 
done in the seminal work of \cite{zafa}, 
whereas for higher dimensional cosets there
is already some work in the literature \cite{BABA}.
A related problem is also finding a consistent quantum theory for 
vortices.
This appears to have been the initial motivation of Lund and Regge 
(see \cite{LuRe}).

%\hfill\newpage

\bs\bs 

%\section*

\centerline{\bf Acknowledgments }
\no
I would like to thank the organizers of the conferences in Imperial college
and in Argonne Nat.Lab. for their warm hospitality and for financial support. 
This work was also carried out with the financial support 
of the European Union Research Program 
``Training and Mobility of Researchers'', under contract ERBFMBICT950362.
It is also work supported by the European Commision TMR program       
ERBFMRX-CT96-0045.

%%%%%%%%%%%%%%%%%%%%%%%

%\section*{References}

\end{document}